\documentclass[zpreprint,zbstdefault]{zeus_paper}
\usepackage[english]{babel}
\newcommand{\ZcoosysA}{%
The ZEUS coordinate system is a right-handed Cartesian system, with the $Z$
axis pointing in the proton beam direction, referred to as the ``forward
direction'', and the $X$ axis pointing left towards the center of HERA.
The coordinate origin is at the nominal interaction point.\xspace}

\newcommand{\ZcoosysfnA}{\footnote{\ZcoosysA}}

\chardef\usc=95
\chardef\til=126
\catcode`\@=11 
\DeclareRobustCommand\xdotspace{\futurelet\@let@token\@xdotspace}
\def\@xdotspace{%
  \ifx\@let@token.\else
  \ifx\@let@token\bgroup.\else
  \ifx\@let@token\egroup.\else
  \ifx\@let@token\/.\else
  \ifx\@let@token\ .\else
  \ifx\@let@token~.\else
  \ifx\@let@token!.\else
  \ifx\@let@token,.\else
  \ifx\@let@token:.\else
  \ifx\@let@token;.\else
  \ifx\@let@token?.\else
  \ifx\@let@token/.\else
  \ifx\@let@token'.\else
  \ifx\@let@token).\else
  \ifx\@let@token-.\else
  \ifx\@let@token\@xobeysp.\else
  \ifx\@let@token\space.\else
  \ifx\@let@token\@sptoken.\else
   .\space
   \fi\fi\fi\fi\fi\fi\fi\fi\fi\fi\fi\fi\fi\fi\fi\fi\fi\fi}
\catcode`\@=12 

\newcommand{\stru}[2]{%
   \relax\ifmmode\hbox{\vrule height#1 depth#2 width0pt}%
   \else\vrule height#1 depth#2 width0pt\fi}

\newcommand{\Ronum}[1]{\uppercase\expandafter{\romannumeral#1}}
\newcommand{\ronum}[1]{\expandafter{\romannumeral#1}}
\DeclareRobustCommand{\LaTeXZ}{%
  \LaTeX\kern-.05em4\kern-.1em
  {\raisebox{-0.2ex}{$\scriptstyle\text{ZEUS}$}}\xspace}

\DeclareMathAlphabet{\mathbf}{OT1}{cmr}{bx}{sl}
\newcommand{\eVdist}{\kern-0.06667em}

\newcommand{\Gev}{{\text{Ge}\eVdist\text{V\/}}}

\newcommand{\gev}{{\,\text{Ge}\eVdist\text{V\/}}}

\newcommand{\Tesla}{\,\text{T}}

\newcommand{\slashfrac}[2]{%
  \raisebox{0.5ex}{\ensuremath #1}\kern-0.12em/\kern-0.08em
  \raisebox{-.8ex}{\ensuremath #2}}

\newcommand{\sqr}[3]{%
    {\vcenter{\hrule height.#3ex\hbox{\vrule width.#2ex height#1ex
     \kern#1ex\vrule width.#3ex}\hrule height.#2ex}}}

\catcode`\@=11 
\newcommand{\parenbar}{\mathpalette\p@renb@r}
\def\p@renb@r#1#2{\vbox{%
  \ifx#1\scriptscriptstyle \dimen@.7em\dimen@ii.2em\else
  \ifx#1\scriptstyle \dimen@.8em\dimen@ii.25em\else
  \dimen@1em\dimen@ii.4em\fi\fi \offinterlineskip
  \ialign{\hfill##\hfill\cr
    \vbox{\hrule width\dimen@ii}\cr
    \noalign{\vskip-.3ex}%
    \hbox to\dimen@{$\mathchar300\hfil\mathchar301$}\cr
    \noalign{\vskip-.3ex}%
    $#1#2$\cr}}}
\catcode`\@=12 

\newcommand{\IP}{{\rm I$\kern-0.01667em$P}\xspace}

\mathchardef\qsm=63
\mathchardef\pls=43
\mathchardef\mns=512
\mathchardef\plm=518
\mathchardef\eql=61
\mathchardef\smallleft=300
\mathchardef\smallright=301
\mathchardef\les=316
\mathchardef\gre=318
\mathchardef\leq=532
\mathchardef\grq=533

\catcode`\@=11 
\newcounter{pict@width}
\newcounter{pict@height}
\newlength{\pict@scale}
\newcommand{\psfigadd}[4]{%
\setcounter{pict@width}{1*\ratio{#2+\pict@scale/2}{\pict@scale}}
\setcounter{pict@height}{1*\ratio{#3+\pict@scale/2}{\pict@scale}}
\setlength{\unitlength}{\pict@scale}
\hbox to #2{\hspace{-\fill}\begin{picture}(\thepict@width,\thepict@height)
\put(0,0){\psfig{figure=#1,width=#2,height=#3,clip=}}
\SetScale{0.283466457}
\SetWidth{1.763889}
{#4}
\end{picture}}
}
\newcounter{pict@widthfst}
\newcounter{pict@widthscd}
\newcounter{pict@widthtot}
\newcommand{\psfigaddtwo}[7]{%
\setcounter{pict@widthfst}{1*\ratio{#2+\pict@scale/2}{\pict@scale}}
\setcounter{pict@widthscd}{1*\ratio{#2+#4+\pict@scale/2}{\pict@scale}}
\setcounter{pict@widthtot}{1*\ratio{#2+#4+#6+\pict@scale/2}{\pict@scale}}
\setcounter{pict@height}{1*\ratio{#3+\pict@scale/2}{\pict@scale}}
\setlength{\unitlength}{\pict@scale}
\hbox{\hspace{-\fill}\begin{picture}(\thepict@widthtot,\thepict@height)
\put(0,0){\psfig{figure=#1,width=#2,height=#3,clip=}}
\put(\thepict@widthscd,0){\psfig{figure=#5,width=#6,height=#3,clip=}}
\SetScale{0.283466457}
\SetWidth{1.763889}
{#7}
\end{picture}}
}
\newcommand{\psfigror}[4]{%
\setcounter{pict@width}{1*\ratio{#2+\pict@scale/2}{\pict@scale}}
\setcounter{pict@height}{1*\ratio{#3+\pict@scale/2}{\pict@scale}}
\setlength{\unitlength}{\pict@scale}
\hbox{\begin{picture}(\thepict@width,\thepict@height)
\put(0,\thepict@height){\psfig{figure=#1,width=#3,height=#2,clip=,angle=270}}
\SetScale{0.283466457}
\SetWidth{1.763889}
{#4}
\end{picture}}
}
\newcommand{\psfigrol}[4]{%
\setcounter{pict@width}{1*\ratio{#2+\pict@scale/2}{\pict@scale}}
\setcounter{pict@height}{1*\ratio{#3+\pict@scale/2}{\pict@scale}}
\setlength{\unitlength}{\pict@scale}
\hbox{\begin{picture}(\thepict@width,\thepict@height)
\put(0,0){\psfig{figure=#1,width=#3,height=#2,clip=,angle=90}}
\SetScale{0.283466457}
\SetWidth{1.763889}
{#4}
\end{picture}}
}
\catcode`\@=12 
\newlength\listtextwidth

\catcode`\@=11 
\newlength{\@tabfninsert}
\newlength{\@tabfnwidth}
\newcommand{\tabfootnote}[2]{%
  \setlength{\@tabfninsert}{0.8em}
  \setlength{\@tabfnwidth}{\textwidth}
  \addtolength{\@tabfnwidth}{-\@tabfninsert}
  \addtolength{\@tabfnwidth}{-0.4em}
  \noindent\makebox[\@tabfninsert][r]{\footnotesize$^{#1}$\hfil}\hfill%
  \parbox[t]{\@tabfnwidth}{\footnotesize #2\hfill}}
\catcode`\@=12 

\newcommand {\pom} {I\!\!P}

\newcommand {\reg} {I\!\!R}

\def\citeCTD{{\cite{%
nim:a279:290,*npps:b32:181,*nim:a338:254%
}}\xspace}
\def\citeCAL{{\cite{%
nim:a309:77,*nim:a309:101,*nim:a321:356,*nim:a336:23%
}}\xspace}

\begin{document}
\prepnum{DESY--04--037}

\title{
Study of the pion trajectory\\
in the photoproduction\\
of leading neutrons at HERA\\
}                                                       
                    
\author{ZEUS Collaboration}
\date{\today}

\abstract{
Energetic neutrons produced
in $ep$ collisions  at HERA have been studied
with the ZEUS detector in the photoproduction regime 
at a mean photon-proton center-of-mass energy of 220~GeV.
The neutrons carry a large fraction $0.64 < x_L <0.925$ of the
incoming proton energy, and the
four-momentum-transfer squared at the proton-neutron vertex  is small,
$|t|<0.425$ GeV$^2$. 
The $x_L$ distribution of the neutrons is measured 
in bins of $t$. 
The $(1-x_L)$ distributions in the $t$ bins studied satisfy a power law 
$dN/dx_L \propto (1-x_L)^{a(t)}$, 
with the powers $a(t)$ following a linear function of
$t$:
\mbox{$a(t)=0.88\pm 0.09 ({\rm stat.})^{+0.34}_{-0.39}({\rm syst.})-(2.81\pm 0.42({\rm stat.})^{+1.13}_{-0.62}({\rm syst.})\ {\rm GeV}^{-2})t$.}
This result is consistent with the expectations of 
pion-exchange models, in which the incoming proton fluctuates
to a neutron-pion state, and the electron interacts with the pion.
}

\makezeustitle

\newcommand{\address}{ }                                                                           
                                                   %
\begin{center}                                                                                     
{                      \Large  The ZEUS Collaboration              }                               
\end{center}                                                                                       
  S.~Chekanov,                                                                                     
  M.~Derrick,                                                                                      
  D.~Krakauer,                                                                                     
  J.H.~Loizides$^{   1}$,                                                                          
  S.~Magill,                                                                                       
  S.~Miglioranzi$^{   1}$,                                                                         
  B.~Musgrave,                                                                                     
  J.~Repond,                                                                                       
  R.~Yoshida\\                                                                                     
 {\it Argonne National Laboratory, Argonne, Illinois 60439-4815}, USA~$^{n}$                       
\par \filbreak                                                                                     
  M.C.K.~Mattingly \\                                                                              
 {\it Andrews University, Berrien Springs, Michigan 49104-0380}, USA
% ...
\newpage                                                                                           
$^{\    1}$ also affiliated with University College London, London, UK \\                          
$^{\    2}$ retired \\                                                                             
% ...
                                                           %
                                                           %
\newpage   
                                                           %
                                                           %
\begin{tabular}[h]{rp{14cm}}                                                                       
$^{a}$ &  supported by the Natural Sciences and Engineering Research                               
          Council of Canada (NSERC) \\                                                             
$^{b}$ &  supported by the German Federal Ministry for Education and                               
          Research (BMBF), under contract numbers HZ1GUA 2, HZ1GUB 0, HZ1PDA 5, HZ1VFA 5\\         
$^{c}$ &  supported by the MINERVA Gesellschaft f\"ur Forschung GmbH, the                          
          Israel Science Foundation, the U.S.-Israel Binational Science                            
          Foundation and the Benozyio Center                                                       
          for High Energy Physics\\                                                                
$^{d}$ &  supported by the German-Israeli Foundation and the Israel Science                        
          Foundation\\                                                                             
$^{e}$ &  supported by the Italian National Institute for Nuclear Physics (INFN) \\                
$^{f}$ &  supported by the Japanese Ministry of Education, Culture,                                
          Sports, Science and Technology (MEXT) and its grants for                                 
          Scientific Research\\                                                                    
$^{g}$ &  supported by the Korean Ministry of Education and Korea Science                          
          and Engineering Foundation\\                                                             
$^{h}$ &  supported by the Netherlands Foundation for Research on Matter (FOM)\\                   
$^{i}$ &  supported by the Polish State Committee for Scientific Research,                         
          grant no. 620/E-77/SPB/DESY/P-03/DZ 117/2003-2005\\                                      
$^{j}$ &  partially supported by the German Federal Ministry for Education                         
          and Research (BMBF)\\                                                                    
$^{k}$ &  supported by RF President grant N 1685.2003.2 for the leading                            
          scientific schools and by the Russian Ministry of Industry, Science                      
          and Technology through its grant for Scientific Research on High                         
          Energy Physics\\                                                                         
$^{l}$ &  supported by the Spanish Ministry of Education and Science                               
          through funds provided by CICYT\\                                                        
$^{m}$ &  supported by the Particle Physics and Astronomy Research Council, UK\\                   
$^{n}$ &  supported by the US Department of Energy\\                                               
$^{o}$ &  supported by the US National Science Foundation\\                                        
$^{p}$ &  supported by the Polish Ministry of Scientific Research and Information                  
          Technology, grant no. 112/E-356/SPUB/DESY/P-03/DZ 116/2003-2005\\                        
$^{q}$ &  supported by the Polish State Committee for Scientific Research,                         
          grant no. 115/E-343/SPUB-M/DESY/P-03/DZ 121/2001-2002, 2 P03B 07022\\                    
\end{tabular}                                                                                      
                                                           %
                                                           %

\newcommand{\xgo}{x_{\gamma}^{\mbox{\rm\tiny OBS}}}
\newcommand{\xpo}{x_p^{\mbox{\rm\tiny OBS}}} 
\newcommand{\xpio}{x_{\pi}^{\mbox{\rm\tiny OBS}}} 
\newcommand{\ETJ}{\ensuremath{E_{\mbox{\rm\tiny T}}^{\mbox{\rm\tiny jet}}}}
\newcommand{\ETAJ}{\ensuremath{\eta^{\mbox{\rm\tiny jet}}}}
\newcommand{\ETAB}{\ensuremath{\bar{\eta}}}
\newcommand{\lap}{\ensuremath{\stackrel{_{\scriptstyle <}}{_{\scriptstyle\sim}}}}
\newcommand{\gap}{\ensuremath{\stackrel{_{\scriptstyle >}}{_{\scriptstyle\sim}}}}

\pagenumbering{arabic} 
\pagestyle{plain}
\section{Introduction}
\label{sec-int}

Several studies
of leading neutron production in $ep$ interactions at HERA 
 have been  reported previously
 \cite{pl:b384:388,np:b596:3,np:b637:3,charm-with-neutron,epj:c6:587}.
Many features of the data are described by
pion exchange models, in which the incoming proton
fluctuates into a neutron-pion state and the pion interacts
with the incoming electron or positron.
The kinematic variables $t$, the square of the four-momentum
transfer at the proton-neutron vertex, and $x_L$,
the energy fraction
of the proton carried by the neutron, are 
convenient variables for studying energy-angle
correlations. 
In this 
study of semi-inclusive photoproduction,
$\gamma p\rightarrow n X$, where the photon is quasi-real, 
the energy distribution
of leading neutrons is measured as a function of $t$, which is  
determined using a new position detector to measure 
the angle of the neutron.
The $(1-x_L)$ distribution is presented as a function of $t$
for large $x_L$ ($0.64<x_L<0.925$) and small $|t|$ ($|t|<0.425$ GeV$^2$).
The results
are interpreted in the context of pion exchange, 
in order to provide a test of the consistency of this 
picture of leading neutron production.

\section{Experimental set-up and kinematics}
\label{sec-exp}

The data used for this measurement were 
collected in the year 2000 at the $ep$ collider HERA with the 
ZEUS detector,
during a short run period in which a special trigger was implemented.
The data set corresponds to an integrated  luminosity  
of 9 pb$^{-1}$. During this period
HERA collided $27.5\gev$ positrons with $920\gev$ protons at a
center-of-mass energy of $318\gev$. 

Charged particles are tracked in the central tracking detector\citeCTD,
which operates in a magnetic field of $1.43\Tesla$ provided by a thin 
superconducting solenoid
and covers the polar-angle{\ZcoosysfnA} 
region 
\mbox{$15^\circ<\theta<164^\circ$}. 
Outside the solenoid is a hermetic, high-resolution 
uranium--scintillator calorimeter (CAL)~\citeCAL used to measure
the energies of the final-state particles.

Bremsstrahlung 
$ep \rightarrow e \gamma p$ and the photoproduction of hadrons
$ep \rightarrow e X$ are tagged using the luminosity (LUMI) detectors
\cite{desy-92-066,*zfp:c63:391,*acpp:b32:2025}. 
The bremsstrahlung photons are measured with a
lead-scintillator calorimeter 
located 107~m from the interaction point in the
positron-beam direction. 
 A similar calorimeter at
35~m from the interaction point is used to measure positrons scattered
at very small angles in an energy range 5-20~GeV, with an
energy resolution of 0.19$\sqrt{E}$ ($E$ in $\Gev$).

\begin{figure} 
\centerline{\epsfig{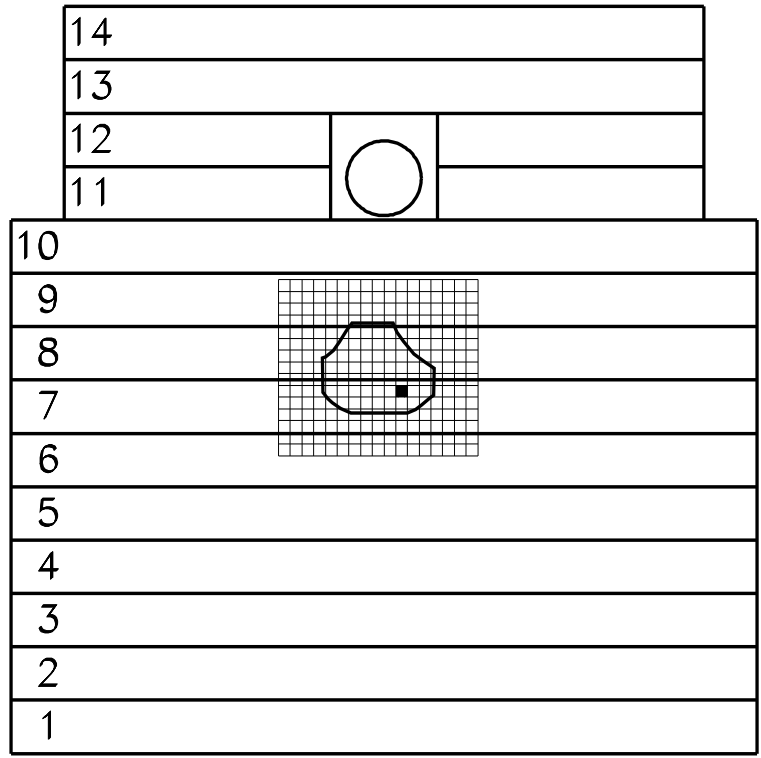}}
\caption{
The FNC as viewed from the rear with the tower numbers indicated on the
left-hand side.
The scintillator strips of the FNT hodoscope are shown 
superposed on towers 6-9
of the FNC.
The irregular contour shows the outline of
the geometric acceptance allowed by the proton beam-line
elements.  
The full square
indicates the approximate position of the projection of the zero-degree 
line.
The proton beampipe, shown as a circle, passes through a hole
in the upper part of the FNC. 
}
\label{fig-PositionFNT}
\end{figure}

ZEUS has two detectors for measuring forward particles,
a leading proton spectrometer (LPS)\cite{zfp:c73:253}, 
consisting of six stations of silicon 
strip detectors, and a forward neutron calorimeter (FNC) 
\cite{nim:a354:479,nim:a394:121,proc:calor97:295}  downstream from
the LPS, 105~m from the interaction point.
The FNC is a
lead-scintillator sampling calorimeter, the front section of which
is divided into 14 towers,  
as illustrated in Fig.~\ref{fig-PositionFNT}.
The hadronic energy resolution,
as measured in a pion test beam at 120~GeV, 
is
$0.65\sqrt{E}$ ($E$ in GeV). 
Three planes of veto counters, 
located 70, 78, and 199 cm 
in front of the calorimeter, are used to detect charged
particles produced by neutrons interacting in inactive material
in front of the FNC. 

Magnet apertures limit the FNC geometric acceptance to neutrons 
inside the area outlined in Fig.~\ref{fig-PositionFNT}. This
limit corresponds to 
neutron production angles less than 
0.8~mrad;
i.e., to transverse momenta $p_T = E_n\theta_n \le 0.74x_L$~GeV. 
The neutron production angle, $\theta_n$,
is measured with respect to the proton beam
direction at the interaction point.
The mean value of $p_T$ of the detected neutrons is 0.22~GeV. 
The overall acceptance of the FNC
is about 25\%
for neutrons with energy \mbox{$x_L > 0.6$}
and \mbox{$\theta_n<0.8$ mrad}.

The forward neutron tracker (FNT), designed to measure the position
of neutron showers, 
was installed in the FNC in 1998.  
It consists of two hodoscope planes of 1.2~cm-wide scintillator strips,
with 17 strips measuring the $X$ position and 15 strips measuring the
 $Y$ position.
Figure~\ref{fig-PositionFNT} shows the position of the FNT hodoscope in the
FNC in the plane transverse to the beam as well as the zero-degree point, the
projection of the proton beam direction at the interaction point onto the FNT.
The FNT is situated approximately one interaction length
inside the FNC. This position is deep enough that a large fraction
of the neutrons begin to shower in front of the FNT, but not so deep
as to compromise the position resolution. 
The position resolution of neutron showers in the FNT
 was measured to be 0.23~cm by placing an adjustable
collimator in front of 
the FNC during special test and calibration runs.


The kinematics of inclusive photoproduction $ep\rightarrow 
e\gamma p \rightarrow eX$ at HERA
is specified by $W$, the photon-proton center-of-mass energy.
This is related to the positron-proton center-of-mass energy, 
$\sqrt{s}$, by $W^2=ys$. The inelasticity, $y$, of the scattered positron
is defined by $y=(E-E')/E$, where $E$~($E'$) is
 the energy of the incoming (scattered) positron.

To describe the neutron-tagged process 
$ep \rightarrow e\gamma p \rightarrow enX$,
the variables $x_L$ and $p_T$ are also used. These are related to $t$ by 
\[
t \approx -\frac{p^2_T}{x_L}-\frac{(1-x_L)}{x_L}\left(m^2_n-x_L m^2_p\right),
\]
where
$m_p$ is the mass of the proton and $m_n$ is the mass of the
neutron.

\section{Event selection}
\label{sec-sel}


The data sample was collected using a trigger that required at least 5 GeV 
in the LUMI electron calorimeter in coincidence with at least 0.5~GeV 
in the rear part 
\mbox{$(\theta > 127^\circ)$}
of the CAL.
 In addition, the trigger required an energy deposit in the FNC corresponding
to $x_L > 0.2$.
The trigger efficiency of the FNC
was close to 100\%
for the $x_L$ range under consideration in this paper
($x_L > 0.64$).

Photoproduction events were selected offline using cuts based
on the reconstructed vertex position and calorimeter energy deposits.
The events were required to have an energy in the
LUMI electron calorimeter in the range $10<E'<18$~GeV
and an energy in the LUMI photon calorimeter smaller than
$1$~GeV to eliminate bremsstrahlung overlap events.
These cuts restricted the photon virtuality, $Q^2$, to values smaller than
$\sim 0.02$~GeV$^{2}$, with a median value
of $\approx 10^{-3}$~GeV$^2$, and 
resulted in a mean value for $W$ of 220 GeV.

Events with a leading neutron were selected 
offline
by requiring  $x_L>0.64$ and by
imposing the following cleaning cuts.
Scattered protons, bent into the top towers 11 to 14 of the FNC (Fig.~1)
by the HERA dipole magnets, 
%
were eliminated 
by requiring that the tower with the maximum energy deposit
be in the range 6 to 9.
Photons were removed by eliminating showers with a small energy-weighted vertical width.

Neutrons that started showering in front of the FNC were
removed by requiring that the scintillator veto counter farthest
from the FNC had an energy deposit below that of a minimum-ionizing particle.
 To minimize false vetoes due to
calorimeter albedo, only the farthest counter was used.

 Finally, only events 
with $x_L < 0.925$ were used in the analysis, since the effect of any
error in the energy scale, and the
influence of exchanges other than the pion, become significant 
 at high $x_L$ (see Sections 5 and 6, respectively).
After these selection  cuts, the data sample consisted of  31\,756
events.

Not all neutrons start to shower in front of the FNT.
To select events with useful position information, it was  required
that the energy deposited in each plane of the FNT be well above pedestal and 
that the strips with the two largest energy deposits be adjacent.
These cuts reduced the sample to 
17\,919 events.

\section{Neutron efficiencies and correction factors}
\label{sec-mc}

The efficiencies and correction factors for the 
leading neutron were calculated with a single-particle Monte Carlo (MC)
simulation. The MC program included the geometry of
the proton beam-line magnets which define the geometric acceptance, 
the details of the absorbing material as obtained from survey measurements, 
the proton beam
divergence, and the measured energy (FNC) and position (FNT) resolutions
for hadronic showers.  The  MC program accounts for the different amount of 
absorbing material in front of the FNC for the two cases when the ZEUS LPS
was, or was not, in the data-taking position.
Events were generated using the observed $x_L$ spectrum and
an exponential distribution in transverse momentum,
\mbox{$e^{-b(x_L)p_T^2}$}.
The slope  $b(x_L)$ was iterated until the MC
distributions matched the
observed uncorrected distributions. The corrected
data distributions were obtained using bin-by-bin unfolding. The two samples of data, corresponding to the two cases 
 when the LPS was, or was not, in the data-taking position, were corrected independently.

The simulation of the absorbing 
material was cross-checked
by producing a material map 
from an analysis of low-energy ($x_L < 0.27$) neutron data, under the
assumption that low $x_L$ neutrons are distributed uniformly over the
geometric acceptance. This assumption is based on noting that
both the maximum $p_T$ accepted by the FNC 
as well as the slope $b$ decrease with $x_L$\cite{np:b637:3},
and hence the geometric acceptance covers
a small region of a slowly changing $p_T$ distribution. 
An analysis using this material map gives results consistent 
  with those obtained using the material
measurements from the survey.

The proton beam had a transverse momentum spread of approximately
0.04 GeV horizontally and 0.1 GeV vertically, as measured
by the LPS.
The variation of
the proton beam orbit was considerably smaller than this spread.
The spread broadens the observed position distribution of the neutrons in 
the FNC, but does not change the peak position. 

\section{Results}
\label{sec-re}

\begin{figure}[htbp!]
\begin{center}\centerline{\epsfig{file=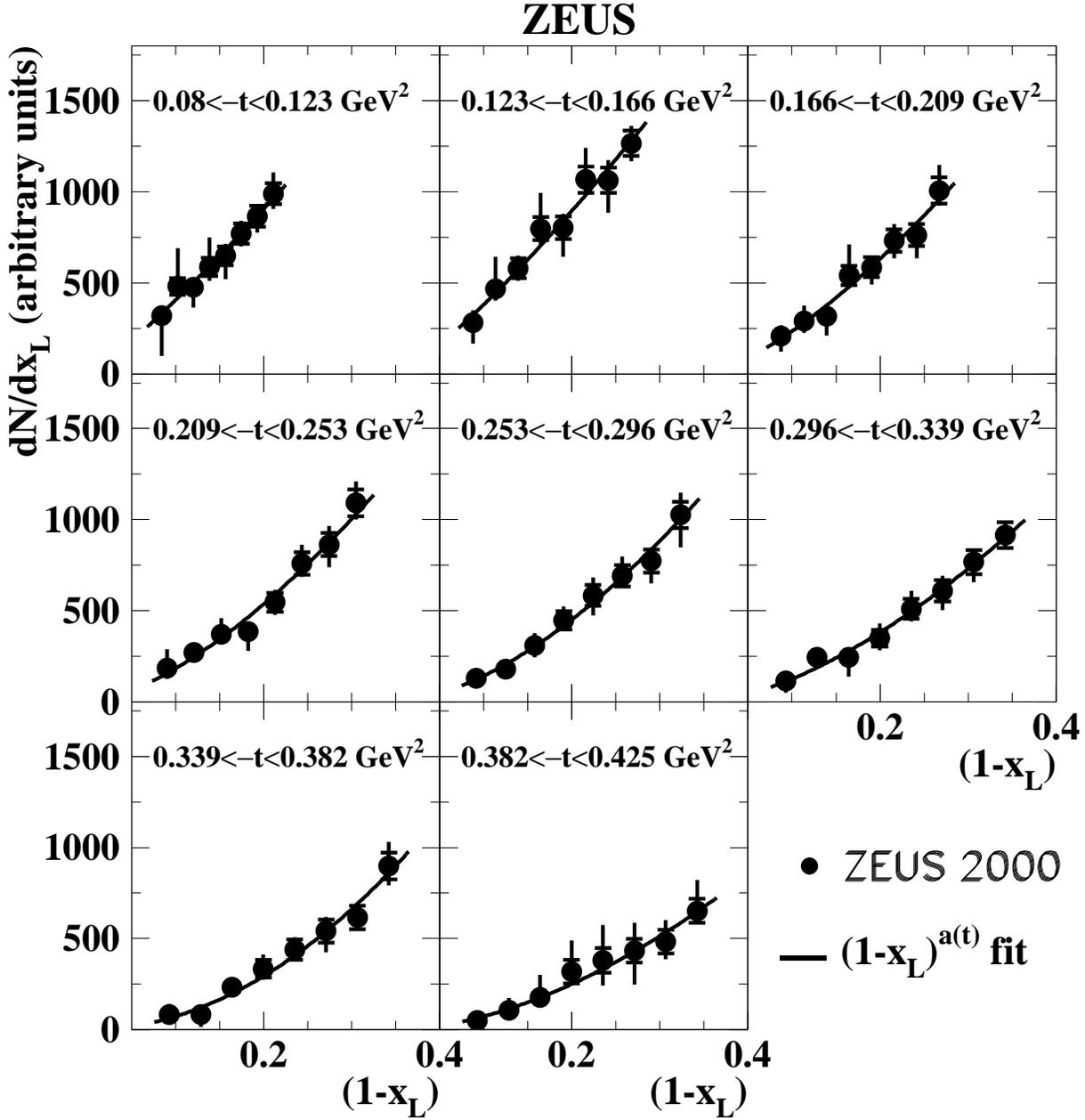,width=16cm}}
\end{center}\caption{
The  $x_L$ distribution
in bins of  $t$. 
The curves show the
results of fits to the data with $0.64<x_L<0.925$
to the form $(1-x_L)^{a(t)}$.
The inner error bars indicate the  statistical uncertainties
and the full bars indicate the quadratic sum of the statistical
and systematic uncertainties.
}
\label{fig-xl}
\end{figure}

The events were binned in $x_L$ and $t$. The bins were chosen to be well within
the acceptance in the $x_L$-$t$ plane and contained 12\,523 events. The 
corrected $x_L$ distributions of leading neutrons 
as a function of $t$ are shown in
Fig.~\ref{fig-xl}.
The distributions are consistent with a power-law dependence in
$(1-x_L)$ of the form $dN/dx_L \propto (1-x_L)^{a(t)}$.
For each $t$ bin the power $a(t)$ was obtained by a 
least-squares fit of this function to the observed distribution.
Only statistical errors were used in the fits because
the systematic uncertainties  are highly correlated.
The resulting fits are superposed on the measured
data points.

There are two main sources of systematic uncertainty: an uncertainty
of  $\pm 2$\% in the absolute energy scale of the 
FNC resulting from the calibration procedure
\cite{proc:calor97:295,np:b637:3,peterf}, 
and an uncertainty of
 $\pm 0.2$ cm in each of the 
$X$ and $Y$ coordinates of the zero-degree point determined from the peak
in the $X,Y$ distribution in the FNT
\cite{thoko}. 
The systematic uncertainties on $a(t)$ were obtained
by varying the energy scale and beam-spot position
within their uncertainties and then repeating the complete 
analysis. The dominant contribution comes from 
the energy-scale uncertainty.

\begin{figure}[htbp!]
\centerline{\epsfig{file=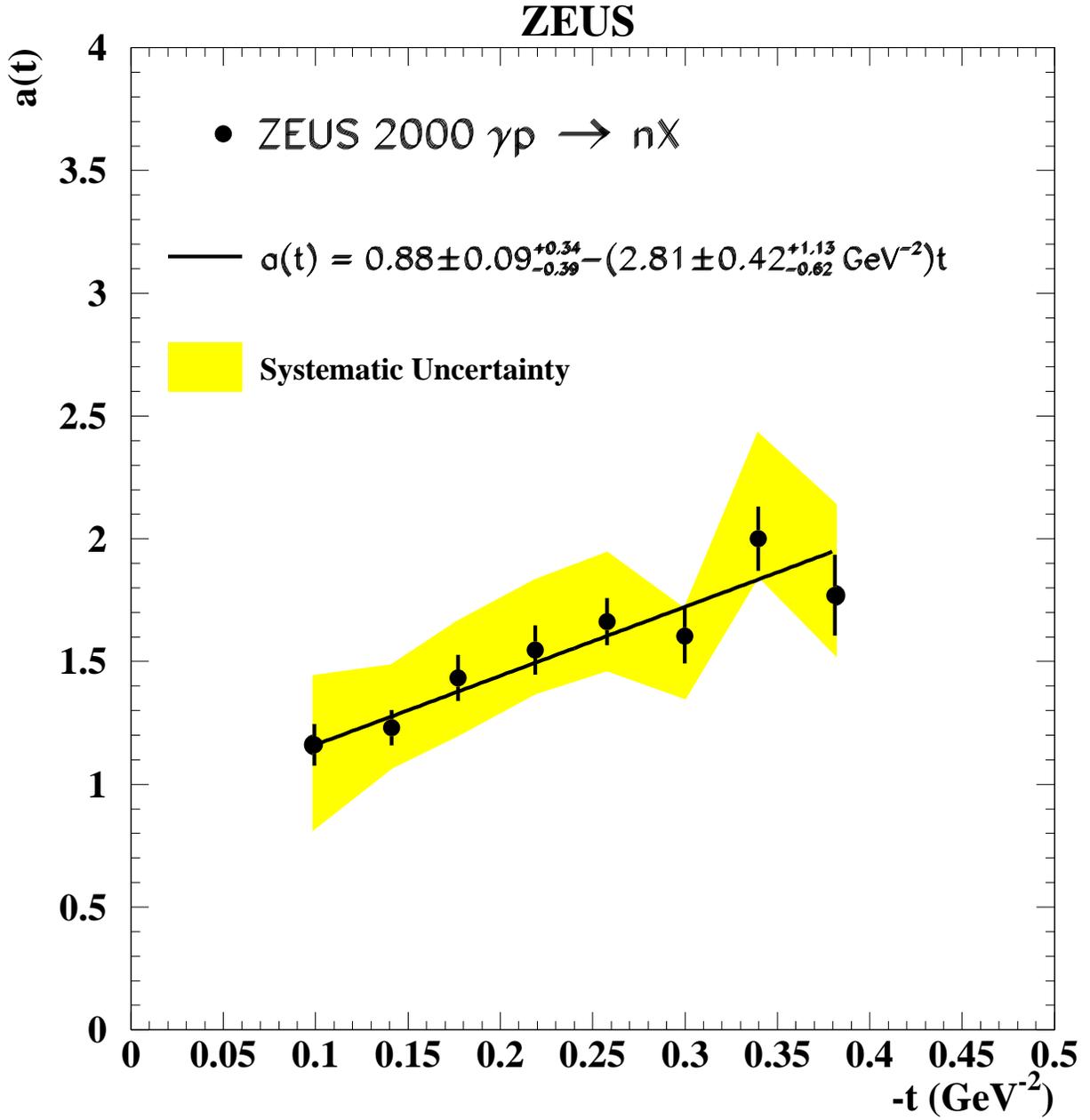,width=16cm}}
\caption{
The powers, $a(t)$, with statistical errors, 
as a function of $-t$.
The solid line is the result of the fit described in the text.
The shaded band 
shows the systematic
uncertainty.
}
\label{fig-3}
\end{figure}

The powers determined from the fits to the data in Fig.~2 are 
plotted as a function of $-t$ in
Fig.~\ref{fig-3}.
They are consistent with a linear function,
\[
a(t)=0.88\pm 0.09 ({\rm stat.})^{+0.34}_{-0.39}({\rm syst.})-(2.81\pm 0.42({\rm stat.})^{+1.13}_{-0.62}({\rm syst.})\ {\rm GeV}^{-2})t.
\]
The correlation between the slope and the intercept is predominantly statistical; the coefficient of  correlation is -0.9.

\section{Discussion}
\label{sec-dis}

Previous experiments ([3]
and references therein)
have shown that leading
neutron production in lepton-hadron and 
hadron-hadron experiments can be described by
pion-exchange models.
The consistency of this description can be tested
by assuming that pion exchange is the dominant mechanism and
deriving the pion Regge trajectory from the measured values
of $a(t)$.

The pion ``flux'', the 
splitting function of a proton to a neutron and pion
($p\rightarrow n \pi^+$), can be written\cite{pl:b38:510} as
\begin{eqnarray}
f_{\pi /p}(x_L ,t) = \frac {1}{4 \pi}\frac{g_{n \pi p}^2}{4 \pi}
\frac{-t}{(m^2_{\pi}-t)^2}(1-x_L )^{1-2 \alpha_ \pi (t)}
\left(F(x_L,t)\right)^2,
\label{eqn-flux}
\end{eqnarray}
where $g_{n \pi p}$ is the coupling at the $n \pi p$ vertex, 
$m_{\pi}$ is the mass of the pion, and
$\alpha_{\pi}(t)=\alpha_{\pi} '( t - m_{\pi}^2)$ is
the Regge trajectory 
of the pion\cite{pitraj1}. 
The function $F(x_L,t)$ is a form-factor which accounts
for the effect of hadronic structure 
on the  $p\rightarrow \pi n$ amplitude
and  for final-state rescattering of the neutron.
As discussed elsewhere\cite{np:b596:3}, this prescription for the flux
with $F(x_L,t)$ set to 1 describes most of the $pn
\rightarrow Xp$ data and also gives a good description of
the ZEUS $x_L$ spectrum for photoproduced
neutrons\cite{np:b596:3,np:b637:3}. In this analysis it is
assumed that $F(x_L,t)$ is a function of $t$ only.

The flux is thus of the form 
$f_{\pi /p}=A(t)\; (1-x_L)^{1-2 \alpha_ \pi (t)}$.
The cross section for neutron production is
given by the product of the flux and 
the total $\gamma \pi$ cross-section 
$\sigma_{\gamma\pi}(s') = 
\sigma_{\gamma\pi}\left( (1-x_L)W^2 \right)  $ as
\[
\frac{d^2 \sigma}{dx_L dt} = A(t) (1-x_L)^{1-2 \alpha_ \pi (t)}
       \sigma_{\gamma\pi}\left( (1-x_L)W^2 \right),
\]
where $s'=(1-x_L)W^2$ is the square of the $\gamma\pi$ center-of-mass energy.
The total $\gamma \pi$ cross section 
is assumed to have
a power law dependence on $s'$, for large $s'$, of
the Donnachie-Landshoff form
\cite{pl:b296:227}
\[
\sigma_{\gamma\pi}(s') = {\cal A}(s')^\epsilon+{\cal B}(s')^{-\eta},
\]
where $\epsilon\approx 0.1$ and $\eta\approx 0.5$.
The two terms correspond to the exchange of the Pomeron ($\pom$)
and the Reggeon ($\reg$), respectively. 
For large $s'$ the Pomeron contribution dominates that of
the Reggeon. In this case the $(1-x_L)$ 
distribution is proportional to $(1-x_L)^{a(t)}$, where
\[
a(t)=1+\epsilon-2\alpha_{\pi} ' t .
\]
Here $m_{\pi}^2$ has been ignored, since its effect on the result
is much smaller than the systematic error.
Therefore, the measured linear function $a(t)$ discussed in Section~5
has an intercept of 
\mbox{$\alpha_{\pom}(0)=(1+\epsilon )$},
the value of the Pomeron trajectory at
\mbox{$t=0$},
and a slope of $2\alpha_{\pi} '$,
twice the slope of the pion trajectory.

It is assumed that the $\pom$ term dominates in the $x_L$ region
of this measurement. The fit\cite{pl:b296:227}
to the total $\gamma p$ cross section
suggests that the highest contribution of the $\reg$ term is
about 12\% at $x_L=0.925$. 
As $x_L$ decreases ($s'$ increases), the Reggeon contribution
falls rapidly. No correction for the Reggeon contribution is applied in
this analysis. 

Previous measurements of the effective Regge trajectory  
in the $p\rightarrow n$ transition
in hadronic interactions\cite{pitraj1,pitraj2,pitraj3}
found it to be dominated by the pion, with 
the intercept in the range -0.1 to 0.3 and the slope
in the range 1.1 to 1.3~GeV$^{-2}$.

The values extracted from the experimental results,
\[
1+\epsilon=\alpha_{\pom}(0)=0.88\pm 0.09 ({\rm stat.})^{+0.34}_{-0.39}({\rm syst.}) 
\] 
and
\[
 \alpha'_{\pi} = 
1.40\pm 0.21({\rm stat.})^{+0.56}_{-0.31}({\rm syst.})\ {\rm GeV}^{-2},
\]
are consistent
with the expectation that
\mbox{$\alpha_{\pom}(0)$} is about 1.1 and
\mbox{$\alpha'_{\pi}$} is about 1~GeV$^{-2}$ 
in the range of $ 0.08 < -t < 0.425 $ 
GeV$^{2}$, and so support the hypothesis that
pion exchange is the dominant process 
in this reaction. Note that the data rule out a significant
role for $\rho$ exchange. If the $\rho$ trajectory,
$\alpha_{\rho} = 0.5 + t$\cite{rhotraj1}, is substituted for the pion
trajectory in Eq.~(1), the  value for $1+\epsilon$ extracted
from the data  is 1.88 rather than 0.88, which is an inconsistent
result for $\alpha_{\pom}(0)$.

\section{Summary}

The dependence of the energy distribution of photoproduced 
leading neutrons on 
the momentum transfer at the proton-neutron vertex
has been studied  
at an average photon-proton center-of-mass energy of 220~GeV.
The $(1-x_L)$ distributions in bins of $t$ are described by a power law,
$dN/dx_L \propto (1-x_L)^{a(t)}$, 
with the powers $a(t)$ following a linear function of
$t$:
\mbox{$a(t)= 0.88\pm 0.09 ({\rm stat.})^{+0.34}_{-0.39}({\rm syst.}) 
- (2.81\pm 0.42({\rm stat.})^{+1.13}_{-0.62}({\rm syst.})\ {\rm GeV}^{-2})t$.}

The linear function can be  interpreted in the framework of 
Regge theory.
The measured values of the intercept and slope are in
agreement with the expectations from pion exchange: i.e.,
the intercept is  the value of the
Pomeron trajectory at $t=0$, and  
the slope is twice the slope of the pion trajectory.
The data thus confirm that the production of 
leading neutrons in photon-proton collisions is well described 
by the pion-exchange model.

\section*{Acknowledgments}

We thank the DESY Directorate for their strong support and encouragement, and 
the HERA machine group for their diligent efforts.  We are grateful for the 
support of the DESY computing and network services.  The design, construction
and installation of the ZEUS detector have been made possible owing to the 
ingenuity and effort of many people who are not
listed as authors.  
This study was only made possible by the physics insight and work of
G.~Levman, to whom we are greatly indebted.

\vfill\eject

{
\def\bibname{\Large\bf References}
\def\refname{\Large\bf References}
\pagestyle{plain}
\ifzeusbst
  \bibliographystyle{./BiBTeX/bst/l4z_default}
\fi
\ifzdrftbst
  \bibliographystyle{./BiBTeX/bst/l4z_draft}
\fi
\ifzbstepj
  \bibliographystyle{./BiBTeX/bst/l4z_epj}
\fi
\ifzbstnp
  \bibliographystyle{./BiBTeX/bst/l4z_np}
\fi
\ifzbstpl
  \bibliographystyle{./BiBTeX/bst/l4z_pl}
\fi
{\raggedright
\bibliography{./BiBTeX/user/syn.bib,%
	      ./BiBTeX/user/extra.bib,%
              ./BiBTeX/bib/l4z_articles.bib,%
              ./BiBTeX/bib/l4z_books.bib,%
              ./BiBTeX/bib/l4z_conferences.bib,%
              ./BiBTeX/bib/l4z_h1.bib,%
              ./BiBTeX/bib/l4z_misc.bib,%
              ./BiBTeX/bib/l4z_old.bib,%
              ./BiBTeX/bib/l4z_preprints.bib,%
              ./BiBTeX/bib/l4z_replaced.bib,%
              ./BiBTeX/bib/l4z_temporary.bib,%
              ./BiBTeX/bib/l4z_zeus.bib}}

\providecommand{\etal}{et al.\xspace}
\providecommand{\coll}{Coll.\xspace}
\catcode`\@=11
\def\@bibitem#1{%
\ifmc@bstsupport
  \mc@iftail{#1}%
    {;\newline\ignorespaces}%
    {\ifmc@first\else.\fi\orig@bibitem{#1}}
  \mc@firstfalse
\else
  \mc@iftail{#1}%
    {\ignorespaces}%
    {\orig@bibitem{#1}}%
\fi}%
\catcode`\@=12
\begin{mcbibliography}{10}

\bibitem{pl:b384:388}
ZEUS \coll, M.~Derrick \etal,
\newblock Phys.\ Lett.{} {\bf B~384},~388~(1995)\relax
\relax
\bibitem{np:b596:3}
ZEUS \coll, J.~Breitweg \etal,
\newblock Nucl.\ Phys.{} {\bf B~596},~3~(2001)\relax
\relax
\bibitem{np:b637:3}
ZEUS \coll, S.~Chekanov \etal,
\newblock Nucl.\ Phys.{} {\bf B~637},~3~(2002)\relax
\relax
\bibitem{charm-with-neutron}
ZEUS \coll, S.~Chekanov \etal,
\newblock Phys.\ Lett.{} {\bf B~590},~143~(2004)\relax
\relax
\bibitem{epj:c6:587}
H1 \coll, C.~Adloff \etal,
\newblock Eur.\ Phys.\ J.{} {\bf C~6},~587~(1999)\relax
\relax
\bibitem{nim:a279:290}
N.~Harnew \etal,
\newblock Nucl.\ Inst.\ Meth.{} {\bf A~279},~290~(1989)\relax
\relax
\bibitem{npps:b32:181}
B.~Foster \etal,
\newblock Nucl.\ Phys.\ Proc.\ Suppl.{} {\bf B~32},~181~(1993)\relax
\relax
\bibitem{nim:a338:254}
B.~Foster \etal,
\newblock Nucl.\ Inst.\ Meth.{} {\bf A~338},~254~(1994)\relax
\relax
\bibitem{nim:a309:77}
M.~Derrick \etal,
\newblock Nucl.\ Inst.\ Meth.{} {\bf A~309},~77~(1991)\relax
\relax
\bibitem{nim:a309:101}
A.~Andresen \etal,
\newblock Nucl.\ Inst.\ Meth.{} {\bf A~309},~101~(1991)\relax
\relax
\bibitem{nim:a321:356}
A.~Caldwell \etal,
\newblock Nucl.\ Inst.\ Meth.{} {\bf A~321},~356~(1992)\relax
\relax
\bibitem{nim:a336:23}
A.~Bernstein \etal,
\newblock Nucl.\ Inst.\ Meth.{} {\bf A~336},~23~(1993)\relax
\relax
\bibitem{desy-92-066}
J.~Andruszk\'ow \etal,
\newblock Preprint \mbox{DESY-92-066}, DESY, 1992\relax
\relax
\bibitem{zfp:c63:391}
ZEUS \coll, M.~Derrick \etal,
\newblock Z.\ Phys.{} {\bf C~63},~391~(1994)\relax
\relax
\bibitem{acpp:b32:2025}
J.~Andruszk\'ow \etal,
\newblock Acta Phys.\ Pol.{} {\bf B~32},~2025~(2001)\relax
\relax
\bibitem{zfp:c73:253}
ZEUS \coll, M.~Derrick \etal,
\newblock Z.\ Phys.{} {\bf C~73},~253~(1997)\relax
\relax
\bibitem{nim:a354:479}
S.~Bhadra \etal,
\newblock Nucl.\ Inst.\ Meth.{} {\bf A~354},~479~(1995)\relax
\relax
\bibitem{nim:a394:121}
S.~Bhadra \etal,
\newblock Nucl.\ Inst.\ Meth.{} {\bf A~394},~121~(1997)\relax
\relax
\bibitem{proc:calor97:295}
S.~Bhadra et al.,
\newblock {\em Proc.\ of the Seventh International Conference on calorimetry in
  High Energy Physics, Tuscon, Arizona, November 1997}, E.~Cheu et al.~(ed.),
  p.~295.
\newblock World Scientific, Singapore (1998)\relax
\relax
\bibitem{peterf}
C.-P.~Fagerstroem.
\newblock Ph.D.\ Thesis, University of Toronto (1999) (unpublished)\relax
\relax
\bibitem{thoko}
T.~Koop.
\newblock Ph.D.\ Thesis, University of Toronto (2004) (unpublished)\relax
\relax
\bibitem{pl:b38:510}
M. Bishari,
\newblock Phys.\ Lett.{} {\bf B~38},~510~(1972)\relax
\relax
\bibitem{pitraj1}
H.~Abramowicz \etal,
\newblock Nucl. Phys.{} {\bf B166},~62~(1980)\relax
\relax
\bibitem{pl:b296:227}
A.~Donnachie and P. V.~Landshoff,
\newblock Phys.\ Lett.{} {\bf B~296},~227~(1992)\relax
\relax
\bibitem{pitraj2}
J.~Hanlon \etal,
\newblock Phys. Rev.{} {\bf D20},~2135~(1979)\relax
\relax
\bibitem{pitraj3}
W.~Flauger and F.~M{\"o}nnig,
\newblock Nucl. Phys.{} {\bf B109},~347~(1976)\relax
\relax
\bibitem{rhotraj1}
R. J. Eden,
\newblock Rep. Prog. Phys.{} {\bf 34},~995~(1971)\relax
\relax
\end{mcbibliography}
}
\vfill\eject



\clearpage

\clearpage

%
%
\end{document}